\documentclass[pdflatex,sn-mathphys-num]{sn-jnl}

\usepackage{graphicx}%
\usepackage{multirow}%
\usepackage{amsmath,amssymb,amsfonts}%
\usepackage{amsthm}%
\usepackage{mathrsfs}%
\usepackage[title]{appendix}%
\usepackage{xcolor}%
\usepackage{textcomp}%
\usepackage{manyfoot}%
\usepackage{booktabs}%
\usepackage{algorithm}%
\usepackage{algorithmicx}%
\usepackage{algpseudocode}%
\usepackage{listings}%
\usepackage{fancyhdr}
\usepackage{lineno}

\newcommand{\dd}{{\rm d}}

\newcommand{\br}{\mathbf{r}}

\newcommand{\bE}{\mathbf{E}}

\newcommand{\sF}{\mathscr{F}}
\newcommand{\sA}{\mathscr{A}}

\providecommand{\norm}[1]{\left\lVert#1 \right\rVert}
\newcommand{\innprod}[2]{\left\langle#1, #2\right\rangle}

\newcommand{\Tr}{\operatorname{Tr}}

\newcommand{\mcI}{\mathcal{I}}

\fancypagestyle{firstpagefooter}
{
    \fancyhf{}
    \fancyfoot[L]{DISTRIBUTION STATEMENT A. Approved for public release: distribution is unlimited.}
}

\unnumbered

\begin{document}
\thispagestyle{firstpagefooter}

\title[Optical measurement is almost quantum: Quantum inspired universal optical bounds]{Optical measurement is almost quantum: Quantum inspired universal optical bounds}

\author*[1]{\fnm{Daniel} \sur{Ish}}\email{daniel.g.ish.civ@us.navy.mil}
\author[1]{\fnm{Chase T.} \sur{Ellis}}

\affil[1]{\orgname{U.S. Naval Research Laboratory}, \orgaddress{\city{Washington}, \state{DC}, \postcode{20375}, \country{USA}}}

\abstract{Inspired by results in quantum information, we report the development of a formalism for optical sensing which abstracts the details of specific optical systems and allows us to form general bounds on measurement based only on the quantity to be measured.
We find a universal bound that controls the performance of any passive optical system regardless of the technology used to construct it and that this bound is attained by a unique measurement scheme for each quantity of interest.
Shockingly, our optical formalism is nearly identical to the quantum formalism, with essentially the same notions of the state to be measured and the operation of the measurement, along with an accompanying uncertainy principle.}


\maketitle

\section{Main}

There is a robust existing literature devoted to determining the limits of optical measurement,\cite{tsangQuantumTheorySuperresolution2016,tsangSemiparametricEstimationIncoherent2019,tylerImagepositionErrorAssociated1982,kolobovQuantumLimitsOptical2000,fellgettAssessmentOpticalImages1997,holmesCramerRaoBounds2013,bornPrinciplesOpticsElectromagnetic2013,barrettFoundationsImageScience2013,prasadInformationoptimizedPhaseDiversity2004} dating back at least to the Rayleigh criterion for the distinguishability of point sources established in 1879.\cite{rayleighXXXIInvestigationsOptics1879}
No one bound covers all optical systems, however: the Rayleigh criterion,\cite{rayleighXXXIInvestigationsOptics1879} the diffraction limit\cite{bornPrinciplesOpticsElectromagnetic2013,barrettFoundationsImageScience2013} and more modern information theoretic\cite{holmesCramerRaoBounds2013,fellgettAssessmentOpticalImages1997,prasadInformationoptimizedPhaseDiversity2004} or quantum bounds\cite{tsangQuantumTheorySuperresolution2016,kolobovQuantumLimitsOptical2000} on optical measurement all use as a starting point a physical model of the optical system under consideration.
This limits the applicability of each bound to the optical system it was developed for, and inhibits comparison among sensing technologies.
Put simply, these bounds answer the question ``how well can this optical system measure this quantity?'' as opposed to ``how well can this quantity be measured by any system?''
This gap becomes even more pressing when accounting for recent progress in inverse designed metamaterial optics, fourier pixels, compressive sensing and algorithmic and optical co-design,\cite{glauserFourierPixelsBidirectional2026,roberts3DpatternedInversedesignedMidinfrared2023,ConformalVolumetricGrayscale,linEndtoendMetasurfaceInverse2022,linEndtoendNanophotonicInverse2021,ballewMultidimensionalWavefrontSensing2023,aryaEndtoEndOptimizationMetasurfaces2024,vargasTimeMultiplexedCodedAperture2021} which perform optical measurement entirely differently from traditional optical systems. These techniques are beyond the purview of existing bounds, and indeed \textit{any} of the typical optical formalisms except for electromagnetism, suggesting we are entering an era where optical designers will enjoy a vastly expanded design space at the cost of dramatically reduced understanding of what designs are desireable. We can contrast this with the situation in quantum information, where, for example, the quantum Cram\'er-Rao bound\cite{helstromMinimumVarianceEstimates1968,sidhuGeometricPerspectiveQuantum2020,braunsteinStatisticalDistanceGeometry1994} provides an ironclad limit on parameter estimation independent of the measurement scheme used for the estimation.
Ultimately this bound is enabled by a clear definition of what consitutes the most general state and measurement scheme for a quantum system.

Here, inspired by the quantum mechanical case, we report the development of a similar formalism for optical sensing which abstracts the details of specific optical systems and allows us to form general bounds on measurement based only on the quantity to be measured.
We find a bound that controls, on equal footing, the performance of any passive optical system, be it a simple imager, an interferometer, a coded aperture system\cite{vargasTimeMultiplexedCodedAperture2021} or more modern approaches like spatial-mode demultiplexing,\cite{tsangQuantumTheorySuperresolution2016} photonic lanterns\cite{birksPhotonicLantern2015} or volumetric metamaterials.\cite{roberts3DpatternedInversedesignedMidinfrared2023,ballewMultidimensionalWavefrontSensing2023}
For each quantity to be estimated, this formalism also identifies a unique measurement scheme which saturates this bound.
Shockingly, this formalism is nearly identical to the quantum formalism, with essentially the same notions of the state to be measured and the operation of the measurement.
The subtle differences will lead to significant impacts, however, with optical measurement exhibiting a stronger uncertainty principle than the quantum one, even preventing optimal measurement of some pairs of commuting observables.
We will also use this bound to report a novel bound on the efficacy of wavefront sensors, together with restrictions on simultaneous attainment of this bound across different wavefront components due to the optical uncertainty principle.

These results offer a new organizing framework for optical design which will allow researchers to pursue more complex, novel sensing schemes and allow apples-to-apples comparison of performance across disparate sensing approaches.
They also point the way towards the development of truly optimal sensing schemes for any given parameter estimation problem in optics.
The formalism developed here also presents promising extensions, including unification with quantum bounds on intensity measurement and analogous limits on hypothesis testing and nonparametric problems, including image formation.

\subsection{Formalism and Bounds}
\begin{figure}
    \centering
    \includegraphics[scale=0.4]{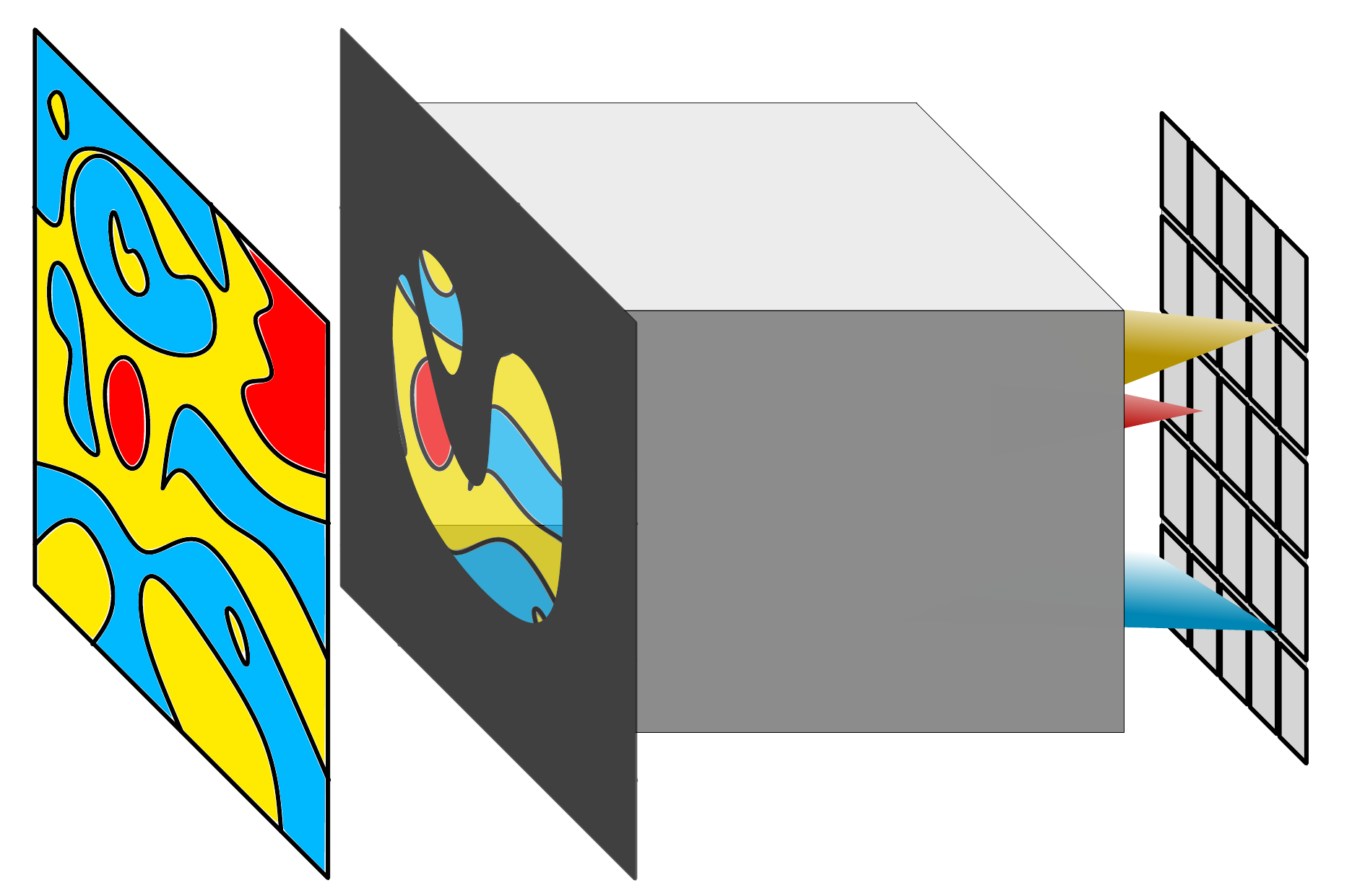}
    \caption{\label{sketch} A schematic depiction of a general optical system.}
\end{figure}
Ultimately, all the systems described above share three features in common: (1) an aperture, defined as the (possibly subdivided) surface which admits light to the system for measurement, (2) finitely many photodetectors, which measure the intensity of the light after manipulation by optical elements and (3) some arrangement of passive, linear optical material which transforms the field pattern at the aperture into the field pattern on the photodetector whose intensity is measured (see Figure~\ref{sketch} for a schematic depiction).
It is this third element that varies so drastically from system to system, so we will abstract it away using only its linearity and the fact that it cannot create optical power not present at the aperture.
Our goal will be a way to write the state of the field and the quantity measured by each pixel in terms of only the degrees of freedom present at the aperture.
To aide the intuition of the reader, we will assume paraxial fields and a simple cannonical form for the detector response (c.f.\cite{barrettFoundationsImageScience2013,loudonQuantumTheoryLight2000,cook1982}).
The interested reader can find discussion of the more general case in the Supplementary Information.

To that end, consider the detected signal in the pixel, given by\cite{barrettFoundationsImageScience2013}
\begin{align}
    X_{ij} = \nu\int_{t_i-\frac{\tau}{2}}^{t_i+\frac{\tau}{2}}\int_{\text{pixel } j}\norm{\bE_F(\br,t)}^2\dd\br\dd t
\end{align}
where $\nu$ represents the quantum efficiency of the focal plane and $\bE_F(\br,t)$ is the time varying electric field at the focal plane surface.
We can regard $\bE_F(\br,t)$ as the member of the Hilbert space, $\sF$, of square integrable functions over the focal plane and time.
With this structure in mind, and defining the operator $\hat{X}_{ij}$ to be multiplication by $\nu$ and the indicator functions of the time interval and area taken up by the pixel, we can recognize that
\begin{align}
    X_{ij} = \innprod{\bE_F}{\hat{X}_{ij}\bE_F}_\sF
\end{align}
with $\innprod{\cdot}{\cdot}_\sF$ the Hilbert space inner product on $\sF$.
Of course, we can also consider the analogous Hilbert space $\sA$ of functions on the aperture.
Using our assumption of linearity, we know that $\bE_F$ is related to the field at the aperture $\bE_A$ by some linear operator $\hat{T}$.
We can use this to write
\begin{align}\label{FPA measurement}
    X_{ij} = \innprod{\bE_A}{\hat{T}^\dagger\hat{X}_{ij}\hat{T}\bE_A}_{\sA}
\end{align}
We have now accomplished our goal of moving all considerations to the degrees of freedom of the aperture, with the measurement being the ``expectation value'' of the operator $\hat{P}_{ij} = \hat{T}^\dagger\hat{X}_{ij}\hat{T}$ in the ``state'' $\bE_A$.
Before we stop to take stock, however, let us take the additional step of allowing the field $\bE_A$ to itself vary stochastically, which allows us to find
\begin{align}\label{measurement equation}
    \overline{X}_{ij} = \Tr\left[\hat{P}_{ij}\hat{\rho}\right]
\end{align}
where $\hat{\rho} = \overline{\bE_A\otimes\bE_A^*}$ with the overline denoting the stochastic average and the tensor product giving a member of $\sA\otimes\sA$. Writing the integral operator for $\hat{\rho}$ allows us to recognize it as the first order optical coherence of the field at the aperture.\cite{barrettFoundationsImageScience2013,loudonQuantumTheoryLight2000}

Equation~\ref{measurement equation} has the suggestive form of the expectation of an operator $\hat{P}_{ij}$ in the state $\hat{\rho}$ in quantum mechanics.
Certainly, the derivation up to this point shows that any information that can be measured by the system must be present in $\hat{\rho}$, so in that sense it must be the proper notion of the state of the field at the aperture.
Using the form of $\hat{\rho}$ as the integral of unnormalized rank 1 projectors, we can see that $\hat{\rho}$ is Hermitian, positive semi-definite and trace class, with trace equal to the average total energy deposited at the aperture.
With the exception of having trace 1, these are the requirements for a density matrix in quantum mechanics.\cite{fanoDescriptionStatesQuantum1957}
So, $\hat{\rho}$ is genuinely an ``unnormalized'' density matrix.

While interesting, this observation is of no practical utility without control over the measurement operators $\hat{P}_{ij}$.
However, even with the limited control we have over $\hat{T}$, we can find that $\hat{P}_{ij}$ is positive semi-definite, because $\hat{X}_{ij}$ is. Furthermore, $\sum_{ij}\hat{P}_{ij}\leq \text{id}_\sA$,  with $\text{id}_\sA$ the identity on $\sA$, since the same is true of the $\hat{X}_{ij}$ and $\hat{T}$ cannot increase the energy deposited on the focal plane above that at the aperture.
Were the inequality an equality (i.e. the system lossless), these operators would therefore form a \textit{positive operator valued measure} (POVM), the most general type of measurement scheme in quantum mechanics,\cite{brandtQuantumMeasurementPositive2003} corresponding to a projective measurement on the state together with auxilary degrees of freedom.\cite{naimark1943representation} So, our description of the measurement performed on that state by an optical system is also identical to the most general description of a measurement performed in quantum mechanics, albeit with the addition of losses.
To cover the slight generalization, we will refer to the set of operators comprising an optical measurement as a \textit{pixel operator measure} (POM).

Inspired by the close correspondence to measurement in quantum mechanics, we can attempt to repeat the argument which leads to the quantum Cram\'er-Rao bound. There, each POVM produces a classical statistical distribution of measurement outcomes, and the variance of any unbiased estimator of a parameter $\theta$ formed from these outcomes is bounded below by the inverse of the Fisher information due to the Cram\'er-Rao bound.\cite{casellaStatisticalInference2024} Maximizing the classical Fisher information over the set of all POVMs then produces a lower bound applicable accross all possible measurements, the quantum Cram\'er-Rao bound.\cite{helstromMinimumVarianceEstimates1968,braunsteinStatisticalDistanceGeometry1994,sidhuGeometricPerspectiveQuantum2020}
It is here that we find the most consequential divergence between the quantum and optical cases.
In the quantum case, the classical statistical distribution represents our probability of observing some set of outcomes, while in the optical case we observe the value of all of these ``probabilities'' at once, since they correspond to the signal in each of our pixels, subject to some noise due to our electronics, background radiation, or uncertainties in the field themselves.
Predictably, this results in a different form for the classical Fisher information.
For this work, we will assume that the dominant noise source is electronics noise which is independent of both $\hat{\rho}$ and $\hat{P}_{ij}$. We model this noise0 by independent, identically distributed Gaussians with mean zero and variance $\sigma^2$. With this, we find the classical Fisher information
\begin{align}
    \mcI_\theta &= \frac{1}{\sigma^2}\sum_{ij}\left(\partial_\theta X_{ij}\right)^2 =\frac{1}{\sigma^2}\sum_{ij}\Tr\left[\hat{P}_{ij}\partial_\theta\hat{\rho}\right]^2
\end{align}
Separating $\partial_\theta\hat{\rho}$ into its positive and negative parts, $\partial_\theta\hat{\rho} = \partial_\theta\hat{\rho}^+ - \partial_\theta\hat{\rho}^-$ and making use of the properties of the POM, we find the \textit{optical Cram\'er-Rao bound}
\begin{align}
    \sigma^2\mcI_\theta \leq \Tr\left[\partial_\theta\hat{\rho}^+\right]^2+\Tr\left[\partial_\theta\hat{\rho}^-\right]^2
\end{align}
with the optimal POM consisting of two projectors $\hat{\pi}_+$ and $\hat{\pi}_-$ which project onto the positive and negative portions of the spectrum of $\partial_\theta\hat{\rho}$.

Both the form of the bound and the optimal measurement scheme are very different from the quantum case.
The latter is perhaps more surprising: in the quantum case,\cite{helstromMinimumVarianceEstimates1968,sidhuGeometricPerspectiveQuantum2020,braunsteinStatisticalDistanceGeometry1994} the optimal measurement consists of projectors onto each distinct eigenspace of the observable of interest, meaning that noncommuting (thus not simultaneously diagonalizable) operators cannot simultaneously be measured perfectly.
Here, since the optimal POM has two members which project onto the \textit{entire} positive and negative spectra, even two $\partial_{\theta_k}\hat{\rho}$ which commute can still fail to simultaneously achieve this bound if one has a positive eigenvalue on an eigenvector for which the other has a negative, in addition to the same restriction on optimal measurement of noncommuting observables.
We can understand this additonal requirement physically through the fact that our observed signal is constrained by photon number conservation, meaning the sum of all intensities is bounded above by the total incident power, while the statistical power is given by the root mean squared signal to noise ratio.
So, intensity devoted to measuring one parameter cannot be devoted to measuring another.

\subsection{A Distant Point Source Seen Through an Aberrating Medium}

As a first application and demonstration of this formalism, we consider the problem of sensing the wavefront of a distant, monochomatic point source seen through a weakly aberrating medium, e.g.\ the guidestar in an adaptive optics system.
For simplicity, we will make the additional assumption that we use only one frame to estimate these parameters, which causes the frame index $j$ to drop out of the analysis and work within scalar diffraction theory\footnote{Of course, the choice to restrict ourselves to scalar diffraction theory means that this bound is not \textit{truly} universal, in that it is only applicable to systems for which scalar diffraction theory captures all relevant physics. The inclusion of polarization significantly complicates the analysis, however, and detracts from our goal of a clean demonstration of the formalism, so will be saved for future work.}.
We can then write the field at the aperture as
\begin{align}
    u(\br|P,\theta) = \sqrt{\frac{P}{A}}\exp\left(i\sum_{k}\theta_k \phi_k(\br)\right)
\end{align}
with $P$ the total power recieved, $A$ the area of the aperture and the numbers $\theta_k$ the aberration components we wish to estimate.
We are free to choose our basis for the aberratons $\phi_k$ as we see fit, so we choose them to integrate to zero and satisfy $\innprod{\phi_l}{\phi_k} = A\delta_{lk}$.
Note the Zernike polynomials satisfy this requirement for a circular aperture when suitably normalized.\cite{bornPrinciplesOpticsElectromagnetic2013}
With this choice, we find that $u(\cdot|1,\theta)$ and $\partial_{\theta_k}u(\cdot|1,\theta)$ form an orthonormal set for fixed values of the aberrations.
We can then write the derivative of the density matrix for the field at the aperture as
\begin{align}
    \partial_{\theta_k}\hat{\rho} = P\left[u(\cdot|1,\theta)\otimes \partial_{\theta_k}u(\cdot|1,\theta)+\partial_{\theta_k}u(\cdot|1,\theta)\otimes u(\cdot|1,\theta)\right]
\end{align}
We can immediately recognize the eigenvectors associated to nonzero eigenvalues as $[u(\cdot|1,\theta)\pm \partial_{\theta_k}u(\cdot|1,\theta)]/\sqrt{2}$ with  eigenvalues $\pm P$.
This gives the bound on the Fisher information
\begin{align}\label{nondimensional}
    \mcI_{\theta_k} \leq 2\left(\frac{P}{\sigma}\right)^2
\end{align}
which does not depend on any $\theta_k$.
Our choice of normalization for $\phi_k$ renders $\theta_k$ nondimensional.
Dimensional factors can be reinserted by expanding the phase due to a parameter of interest, e.g. angle of incidence $\varphi$, in the aberrations $\phi_k$.
This results in, for example, the lower bound on the the noise in any estimator of the angle of incidence at normal incidence $\sigma_{\varphi}\geq \frac{\sqrt{2}}{\pi}\frac{\lambda}{D}\frac{\sigma}{P}$, with $\lambda$ the wavelength of the incident light and $D$ the diameter of the aperture (assumed to be circular).
This bound can be compared to the classic result for a quadrant detector\cite{tylerImagepositionErrorAssociated1982}, $\sigma_\varphi = \frac{3\pi}{16} \frac{\lambda}{D}\frac{\sigma}{P}$ to find that quadrant detectors are shockingly close to optimal for measuring tilt, with noise only equal to approximately 130\% of the bound derived here.

There is one subtlety, in that the careful reader of Reference~\cite{tylerImagepositionErrorAssociated1982} will note that the noise term in that calculation is actually the noise due to both the pixels on either side of the quadrant detector, i.e. $\sqrt{2}$ times the noise of a single pixel.
So, the scheme so near to the bound should actually be thought of as a spot focused on the edge between two large pixels, rather than the corner between four, thus measures \textit{only} tilt and not tip.
If we want to measure both angles of incidence using a scheme of this form, we must sacrifice our accuracy in both by $\sqrt{2}$.
Returning to our analysis, we can see that $\partial_{\theta_k}\hat{\rho}$ does not commute with $\partial_{\theta_l}\hat{\rho}$ for $l\neq k$.
So, this trade off is not an artifact of the particular measurement scheme implemented by the quadrant detector but rather a fundamental limitation: both angles of incidence cannot be simultaneously optimally measured.
Indeed, no two aberration components can be simultaneously optimally measured.
More broadly, note also that $\partial_{\theta_k}\hat{\rho}(\theta)$ also does not generically commute with itself at distinct values of the aberration, so that optimal measurements of aberrations for all possible values of the aberration are impossible.
In particular, one cannot build a system which senses the direction of a point source optimally for all incident angles.

\section{Discussion}

Though we believe that this formalism has great promise as a unifying picture for optical measurement, ultimately this work points to siginificant additional work that is needed to fully develop it.
As a starting point, the existing menagerie of optical systems needs to be analyzed within this framework to serve as a baseline of comparison for future development.

More fundamentally, saturating the optical Cram\'er-Rao bound offers \textit{local} optimality, and there are two questions of a more global nature that remain open.
First, within the context of parametric estimation, there is the question of system performance over extended sets of the parameter to be estimated.
In the wavefront context, this is the question of balancing performance over extended regions of the aberrations, which we know cannot be simultaneously measured optimally.
There's also the nonparametric question: what does this physics imply about limitations on image formation and resolution?

Though the fact of the optical uncertainty principle is remarkable and unexpected, it's current form leaves much to be desired.
As an example, one might ask whether the $\sqrt{2}$ cost of simultaneous measurement of both angles of inicidence displayed by the quadrant detector can be strictly improved, which the present formalism does not address.
Conceptual work can be shared here with the quantum information literature, which to our knowledge does not address optimal estimation two parameters which result in noncommuting observables using an arbitrary POVM, though some work seems to present ingredients for an answer.\cite{leeUncertaintyRelationErrors2020}

This leaves perhaps the most pressing open question raised by this work: why does this correspondence between optical and quantum measurement exist?
This finding is perhaps especially surprising given the seeming uniqueness of quantum measurement as a physical process.
In general, we can see that formally the root of this correspondence is Equation~\ref{FPA measurement}, which is physically the statement that the photodetector is only sensitive to intensity and not to overall phase.
One possible interpretation is that optical and quantum measurement are two examples of a broader class of measurement, call it ``incoherent measurement,'' which arises when a degree of freedom carrying phase is measured in a manner insensitive to that phase.
In principle, one might find other examples of such measurements.
On the other hand, however, it should be noted that the physical process underlying this measurement is itself quantum mechanical in nature, so it may also be the case that this phenomenon is a macroscopic manifestation of the underlying quantum measurement process of photodetection.

\backmatter

\bmhead{Supplementary information}
Supplementary Information is available for this paper

\bmhead{Acknowledgements}
D. Ish and C. Ellis acknowledge support from the Office of the U.S. Undersecretary of Defense for Research and Engineering.
The authors would like to thank Christopher White, Zachary Dromsky, Michael Swift, Sarah Del Ciello and Mitesh Amin for helpful discussions.

\bmhead{Author Contributions}
D. Ish concieved of the approach, performed all pen and paper calculations, and wrote the publication.
C. Ellis assisted in the conceptualization of the work and produced simulations not reported here which drove development of the formalism.

\bmhead{Competing Interests}
The authors declare no competing interests

\bmhead{Correspondence} Correspondence and requests for materials should be addressed to D. Ish.




\end{document}


\thispagestyle{firstpagefooter}
\title[Supplementary Information for ``Optical measurement is almost quantum: Quantum inspired universal optical bounds'']{Supplementary Information for ``Optical measurement is almost quantum: Quantum inspired universal optical bounds''}

\author*[1]{\fnm{Daniel} \sur{Ish}}\email{daniel.g.ish.civ@us.navy.mil}
\author[1]{\fnm{Chase T.} \sur{Ellis}}

\affil[1]{\orgname{U.S. Naval Research Laboratory}, \orgaddress{\city{Washington}, \state{DC}, \postcode{20375}, \country{USA}}}

\maketitle

\section*{Formalism and Bounds}
In this section, we repeat the derivation of the pixel operator measure formalism and optical Cram\'er-Rao bound in more detail with additional commentary for the interested reader.
We caution the reader that we will not be treating in detail the more mathematical questions that arise from this derivation.
\subsection*{Pixel Operator Measures}
Our starting point is the same, with the signal in the $j$th pixel during the $i$th integration time given by
\begin{align}\label{intensity}
    X_{ij} = \nu\int_{t_i-\frac{\tau}{2}}^{t_i+\frac{\tau}{2}}\int_{\text{pixel } j}\norm{\bE_F(\br,t)}^2\dd\br\dd t
\end{align}
Note that we have implictly chosen a ``natural'' unit system to avoid carrying conversion factors between field intensity and the collected charge in the focal plane, as these factors are unimportant for our analysis.
We can define
\begin{align}
    \chi_{ij}(\br,t) = \left\{\begin{array}{lr}
        \nu & \text{ if }\br \text{ is in the $j$th pixel and }t_i-\frac{\tau}{2}\leq t\leq t_i+\frac{\tau}{2}\\
        0 & \text{else}
    \end{array}\right.
\end{align}
to find
\begin{align}
    X_{ij} &= \int_{-\infty}^\infty\int_F\chi_{ij}(\br,t)\norm{\bE_F(\br,t)}^2\dd\br\dd t\nonumber\\
    &  = \int_{-\infty}^\infty\int_F\bE^*(\br,t)\cdot\left[\chi_{ij}(\br,t)\bE_F(\br,t)\right]\dd\br\dd t
\end{align}
with $F$ the whole domain occupied by the focal plane.
Defining the operator
\begin{align}
    \left[\hat{X}_{ij}\bF\right](\br,t) = \chi_{ij}(\br,t)\bF(\br,t)
\end{align}
and recognizing the $L^2$ inner product
\begin{align}
    \innprod{\bA}{\bB}_{\sF} = \int_{-\infty}^\infty\int_F\bA^*(\br,t)\cdot\bB(\br,t)\dd\br\dd t
\end{align}
this gets us
\begin{align}\label{general}
    X_{ij} = \innprod{\bE_F}{\hat{X}_{ij}\bE_F}_\sF
\end{align}
where $\sF$ is the Hilbert space of square integrable functions over $F\times(-\infty,\infty)$.
Using the linearity of the materials which compose the system, there is some linear $\hat{T}$ such that $\bE_F = \hat{T}\bE_A$.
So,
\begin{align}
    X_{ij} &= \innprod{\hat{T}\bE_A}{\hat{X}_{ij}\hat{T}\bE_A}_\sF\\
    & = \innprod{\bE_A}{\hat{T}^\dagger\hat{X}_{ij}\hat{T}\bE_A}_\sA\\
    & = \Tr_\sA\left[\hat{T}^\dagger\hat{X}_{ij}\hat{T}\left(\bE_A\otimes\bE_A^*\right)\right]
\end{align}
where $\sA$ is the Hilbert space of square integrable functions over $A\times(-\infty,\infty)$, with $A$ the domain of the aperture.
In the last line we've used a standard manipulation from quantum mechanics and ``inserted a resolution of the identity.''
It may be more familiar in bra-ket notation:
\begin{align}
    \inoprod{\alpha}{\hat{A}}{\beta} = \sum_i \inprod{\alpha}{i}\inoprod{i}{\hat{A}}{\beta} = \sum_i\inoprod{i}{\hat{A}}{\beta}\inprod{\alpha}{i} = \Tr\left[\hat{A}\ket{\beta}\bra{\alpha}\right]
\end{align}
Now, under suitable assumptions, we can simply let the stochastic average pass through the trace to recieve
\begin{align}
    \overline{X}_{ij} &= \Tr\left[\hat{T}^\dagger\hat{X}_{ij}\hat{T}\left(\overline{\bE_A\otimes\bE_A^*}\right)\right]\\
    & = \Tr\left[\hat{P}_{ij}\hat{\rho}\right]
\end{align}
with $\hat{P}_{ij} = \hat{T}^\dagger\hat{X}_{ij}\hat{T}$ and $\hat{\rho} = \overline{\bE_A\otimes\bE_A^*}$.

Now we turn our attention to establishing the properties of $\hat{P}_{ij}$ and $\hat{\rho}$.
$\hat{\rho}$ is positive semi-definite because
\begin{align}
    \innprod{\bA}{\bE_{A}\otimes\bE_A^*\bA} = \abs{\innprod{\bE_A}{\bA}}^2 \geq 0
\end{align}
so that
\begin{align}
    \innprod{\bA}{\hat{\rho}\bA} = \overline{\abs{\innprod{\bE_A}{\bA}}^2 }\geq 0
\end{align}
Similarly, Hermiticity of $\hat{\rho}$ follows from that of $\bE_{A}\otimes\bE_A^*$.
Finally, we can find
\begin{align}
    \Tr\left[\hat{\rho}\right] = \overline{\Tr\left[\bE_{A}\otimes\bE_A^*\right]}= \overline{\norm{\bE_A}_\sA^2}\label{energy}
\end{align}
where $\norm{\cdot}_\sA$ denotes the $L^2$ norm on $\sA$, justifying that the trace of $\hat{\rho}$ is the average energy depositited at the aperture since
\begin{align}
    \norm{\bE_A}_\sA^2 = \int_{-\infty}^\infty\int_A \norm{\bE_A(\br,t)}^2\dd\br\dd t
\end{align}
One can see that $\hat{\rho}$ is the first order coherance by writing the integral kernel for it (with the tensor product in $\bbR^3\otimes \bbR^3$)
\begin{align}
    \hat{\rho}(\br_2,t_2;\br_1,t_1) = \overline{\bE_A(\br_2,t_2) \otimes\bE^*_A(\br_1,t_1)}
\end{align}

As for $\hat{P}_{ij}$, we can see that since $\hat{X}_{ij}$ is positve semi-definite and Hermitian, so too must $\hat{P}_{ij}$ be.
Since we have considered functions over $F$ only rather than the whole of $\bbR^3$,
\begin{align}
    \sum_{ij}\hat{X}_{ij} = \nu \text{id}_\sF
\end{align}
which implies
\begin{align}
    \sum_{ij}\hat{P}_{ij} = \nu\hat{T}^\dagger\hat{T}
\end{align}
On physical grounds, we must have that
\begin{align}
    \norm{\hat{T}}_{\text{op}}\leq 1
\end{align}
where $\norm{\cdot}_{\text{op}}$ is the operator norm.
Were this not true we would have some $\bE_A$ such that
\begin{align}
    \norm{\hat{T}\bE_A}_\sF>\norm{\bE_A}_\sA
\end{align}
indicating that more energy was deposited on the focal plane than at the aperture.
So, this implies $\norm{\hat{T}^\dagger\hat{T}}_{\text{op}}<1$ and 
\begin{align}
    \sum_{ij}\hat{P}_{ij} \leq \text{id}_\sA
\end{align}
since the efficiency must also have $\nu \leq 1$.

The savvy reader might object at several points of this argument, particularly at Equation~\ref{intensity} as an oversimplification of detector response and the discussion surrounding Equation~\ref{energy} as eliding some of the details of energy transport in electromagnetism.
Ultimately, this presentation is true only in a particular limit, when the plane wave expansion of the fields both at the aperture and at the detector predominantly contain weight with fields close to normal incidence and the frequency spectrum is narrowly peaked around some particular frequency.
More generally, the right starting point is an assumption that the signal recieved by the detector is proportional to the number of photons absorbed in the detector.
Though the details of the operators in question become complicated and detector-specific, the signal in the detector will still be some bilinear in the fields and photon number conservation guarentees that\cite{loudonQuantumTheoryLight2000,cook1982,barrettFoundationsImageScience2013}
\begin{align}
    &X_{ij} = \innprod{\bE_F}{\hat{X}_{ij}\bE_F}\\
    &\sum_{ij}\hat{X}_{ij}\leq \hat{\Phi}_F
\end{align}
where $\hat{\Phi}_F$ is the operator which counts the total number of photons which pass the domain $F$ over the whole time history (accounting both for the photon energy and the obliquity of the field in the plane wave expansion).
On physical grounds, $\hat{X}_{ij}$ must be Hermitian and positive definite and we must have
\begin{align}
    \hat{T}^\dagger\hat{\Phi}_F\hat{T}\leq \hat{\Phi}_A
\end{align}
due to the condition that our system does not create energy, so that the actual bound for the POM reads in general
\begin{align}
    \sum_{ij}\hat{P}_{ij}\leq \hat{\Phi}_A
\end{align}
That is, a photon must pass through the aperture to be measured.
The accompanying optical Cram\'{e}r-Rao bound then must also change to 
\begin{align}
    \sigma^2\mcI_\theta \leq \Tr\left[\hat{\Phi}_A\partial_\theta\hat{\rho}^+\right]^2+\Tr\left[\hat{\Phi}_A\partial_\theta\hat{\rho}^-\right]^2
\end{align}
Due to the need to carefully formulate the operators in question starting from quantum electrodynamics, (c.f. \cite{cook1982}) the details of making this argument rigorous will be left for future work. For now we content ourselves only with having shown the bound in the paraxial quasimonochromatic case.

For the second section of the paper, we drop the frame index $i$ and time dependence.
Ultimately, we anticipate that version of the formalism presented here will be more useful in routine calculations, but arriving at it actually involves a subtle dance of timescales.
For the sake of concreteness, let us assume that $\hat{T}$ is given by some integral kernel and exhibits time translation invariance, though we won't explore conditions under which this is true in this work and expect that these considerations can be formulated without this assumption at the cost of some additional abstraction.
In this case, with $\bT$ a matrix valued integral kernel
\begin{align}
    \bE_F(\br,t) = \int_{-\infty}^\infty\int_A \bT(\br,\br^\prime;t - t^\prime)\bE_A(\br^\prime,t^\prime)\dd\br^\prime\dd t
\end{align}
Using the cannonical form of the signal in Equation~\ref{intensity}, we can see that $\hat{P}_{ij}$ is given by integration against the matrix valued kernel $\bP_{ij}$
\begin{align}
    \bP_{ij}(\br_1,t_1;\br_2,t_2) = \int_{-\infty}^{\infty}\int_A \bT^\dagger(\br,\br_1;t-t_1)\chi_{ij}(\br,t)\bT(\br,\br_2;t-t_2)\dd\br\dd t
\end{align}
Writing $\chi_j(\br)$ for the indicator function of the $j$th pixel and using the temporal Fourier transform, this becomes
\begin{align}
    \bP_{ij}(\br_1,t_1;\br_2,t_2) = \tau\int_A \int_{-\infty}^{-\infty}\int_{-\infty}^{-\infty}\chi_j(\br)\bT^\dagger(\br,\br_1;\omega_1)\bT(\br,\br_2;\omega_2)e^{-2\pi i\omega_1t_1 + 2\pi i\omega_2 t_2}e^{2 \pi i\omega_s t_i}\sinc\left(\omega_s\tau\right)\dd\omega_1\dd\omega_2\dd\br
\end{align}
with the definition of the slow frequency $\omega_s =  \omega_1 - \omega_2$. Additionally defining the fast frequency $\omega_f = (\omega_1+\omega_2)/2$ and the fast and slow times $t_f = t_1 - t_2$ and $t_s= (t_1+t_2)/2$, we find
\begin{align}
    \bP_{ij}(\br_1,t_1;\br_2,t_2) &= \int_A \int_{-\infty}^{-\infty}\int_{-\infty}^{-\infty}\chi_j(\br)\bT^\dagger(\br,\br_1;\omega_f+\omega_s)\bT(\br,\br_2;\omega_f - \omega_s)\tau\sinc\left(\omega_s\tau\right)\nonumber\\
    &\hspace{190pt}\times e^{2\pi i\omega_s t_s + 2\pi i\omega_f t_f}e^{2 \pi i\omega_s t_i}\dd\omega_s\dd\omega_f\dd\br
\end{align}
Now, if we define
\begin{align}
    \rho(\br_1,\br_2| \omega_f,\omega_s) = \int_{-\infty}^{\infty}\int_{-\infty}^\infty \overline{\bE_A(\br_1,t_s+t_f)\otimes\bE_A(\br_2,t_s-t_f)}e^{2\pi i\omega_f t_f + 2\pi i \omega_s t_s}\dd t_f\dd t_s
\end{align}
we get
\begin{align}\label{fourier}
    \overline{X}_{ij} &= \int_A \int_A \int_A \int_{-\infty}^{-\infty}\int_{-\infty}^{-\infty}\chi_j(\br)\bT^\dagger(\br,\br_1;\omega_f+\omega_s)\bT(\br,\br_2;\omega_f - \omega_s)\tau\sinc\left(\omega_s\tau\right)\nonumber\\
    &\hspace{220pt}\times\rho(\br_1,\br_2| \omega_f,\omega_s)e^{2 \pi i\omega_s t_i}\dd\omega_s\dd\omega_f\dd\br\dd\br_1\dd\br_2
\end{align}

Notice that the effect of the integration time is to produce a low pass filter on the slow frequency, $\sinc(\omega_s\tau)$.
If we imagine that this low pass is sharp enough that the product of the two transfer matrices is approximately constant the range of $\omega_s$ not attenuated by this filter, we recieve the form
\begin{align}
    \bP_{ij}(\br_1,t_1;\br_2,t_2) \approx \chi_i(t_s)\int_{-\infty}^{-\infty} \bP_j(\br_1,\br_2;\omega_f)e^{2\pi i\omega_f t_f}\dd\omega_f
\end{align}
with $\chi_i(t)$ the indicator function of the $i$th integration time and
\begin{align}
    \bP_j(\br_1,\br_2|\omega_f) = \int_A \chi_j(\br)\bT^\dagger(\br,\br_1;\omega_f)\bT(\br,\br_2;\omega_f)\dd\br
\end{align}
If we define (with the tensor product in $\bbR^3\otimes\bbR^3$)
\begin{align}
    \overline{\rho}(\br_1,\br_2| \omega_f,t_i) = \int_{t_i-\frac{\tau}{2}}^{t_i+\frac{\tau}{2}}\int_{-\infty}^\infty \overline{\bE_A(\br_1,t_s+t_f)\otimes\bE_A(\br_2,t_s-t_f)}e^{2\pi i\omega_f t_f}\dd t_f\dd t_s
\end{align}
this means
\begin{align}\label{freq}
    \overline{X}_{ij} \approx \int_{-\infty}^\infty \Tr_{\overline{\sA}}\left[\hat{P}_j(\omega_f)\hat{\overline{\rho}}(\omega_f,t_i)\right]\dd \omega_f
\end{align}
where $\overline{\sA}$ is the Hilbert space of functions over the aperture, \textit{without} time.
We will call this the \textit{quasistationary} form.
This, of course, is the form that's more intuitive when considering a typical optical system, where each frequency of incoming light independently passes through the system according to its own transfer function and can think of our signal as having some spectral content that varies slowly in time.
Apparently, for this to be true regardless of the properties of the light in question requires the product of the two transfer functions to depend weakly enough on frequency.
Exploring this condition in full generality is well outside the scope of this work, but we note in passing that Chapter 9 of Reference~\cite{barrettFoundationsImageScience2013} contains an argument that systems which create a focal spot satisfy this condition, along with discussions of similar coherance questions in a more traditional imaging context.
Conversely, one can imagine deliberately creating a system with narrow geometric resonances, e.g. a Fabry-Perot cavity, to increase sensitivity to coherance effects.

The form in Equation~\ref{freq} can also be obtained from Equation~\ref{fourier} if the first order coherance is itself narrowly peaked in $\omega_s$.
Physically, this corresponds (equivalently) to the conditions that the correlations of the field are slowly varying as a function of the slow time or that correlations between disctinct frequencies are small.
This is satisfied in particular by monochomatic sources, justifying the analysis in the second section of the main text.
Additionally, if $\bE_A$ is ergodic,
\begin{align}
    \int_{t_i-\frac{\tau}{2}}^{t_i+\frac{\tau}{2}}\int_{-\infty}^\infty \bE_A(\br_1,t_s+t_f)\otimes\bE_A(\br_2,t_s-t_f)e^{2\pi i\omega_f t_f}\dd t_f\dd t_s \approx \overline{\rho}(\br_1,\br_2| \omega_f,t_i) 
\end{align}
for $\tau$ sufficiently long, meaning that we should expect to measure the stochastic average pixel signal for ergodic sources.

From the point of view of view of the optical Cram\'er-Rao bound and the optimal POM for a given parametric estimation problem, it's actually the condition on the first order coherance which is determinative of whether the quasistationary form is more appropriate.
If the first order coherance contains significant frequency off-diagonal correlations and these correlations vary as a function of the parameter to be estimated, the optimal POM will exploit these correlations and result in pixel operators for which the quasistationary form is not a good approximation.
Put another way, if there is meaningful variation in $\rho$ as a function of the slow time which depends on the estimand $\theta$, $\partial_\theta\hat{\rho}$ will also have such variations and the optimal POM should be thought of collecting the full time histories associated with the positive or negative eigenspaces of these operators into just two measurements, requiring engineering of $\hat{T}$ at all frequencies simultaneously.
Physically, we might imagine estimating the duration or shape of an ultrafast pulse falling into this category.
Conversely, for monochromatic or stationary, incoherent $\hat{\rho}$, the quasistationary formalism will capture all meaningful variation.

\subsection*{Local Estimation and the Optical Cram\'er-Rao Bound}
Consider some parameter $\theta$ on which the fields at the aperture depend.
Our goal is to use the values that the pixels take to form an estimate of $\theta$.
We will restrict our attention to the case where the dominant noise source is the detector itself and model this noise as independent identically distributed additive Gaussian errors with variance $\sigma^2$.
We will also keep the index of the POM general, simply writing $i$, since the precise structure of this index is not critical.
The Fisher information for a fixed POM is then the well-known form for a Gaussian family
\begin{align}
    \mcI_{\theta} =\frac{1}{\sigma^2}\sum_i \left(\partial_\theta X_i(\theta)\right)^2 = \frac{1}{\sigma^2}\sum_i \Tr\left[\hat{P}_i\partial_\theta\hat{\rho}\right]^2
\end{align}
Define $\partial_\theta\hat{\rho}^\pm$ positive semi-definite so that $\partial_\theta\hat{\rho} = \partial_\theta\hat{\rho}^+ - \partial_\theta\hat{\rho}^-$ and compute
\begin{align}
    \sigma^2\mcI_{\theta} &= \sum_i \Tr\left[\hat{P}_i\partial_\theta\hat{\rho}\right]^2\label{start}\\
    &\leq \sum_{i = 1}^N\Tr\left[\hat{P}_i\partial_\theta\hat{\rho}^+\right]^2 + \Tr\left[\hat{P}_i\partial_\theta\hat{\rho}^-\right]^2\\
    &\leq \Tr\left[\left(\sum_i\hat{P}_i\right)\partial_\theta\hat{\rho}^+\right]^2+\Tr\left[\left(\sum_i\hat{P}_i\right)\partial_\theta\hat{\rho}^-\right]^2\\
    &\leq\Tr\left[\partial_\theta\hat{\rho}^+\right]^2+\Tr\left[\partial_\theta\hat{\rho}^-\right]^2
\end{align}
where we have used throughout that $\Tr\left[PQ\right]\geq 0$ if $P\geq 0$ and $Q\geq 0$.
The first inequality follows from neglecting negative cross terms, the second from the fact that $\norm{\bx}_2\leq \norm{\bx}_1$ and the third from the sum property of POMs.
The first inequality is saturated if and only if either $\Tr\left[\hat{P}_i\partial_\theta\hat{\rho}^+\right] = 0$ or $\Tr\left[\hat{P}_i\partial_\theta\hat{\rho}^-\right] = 0$ for each $i$, and the second inequality is saturated if and only if exactly one of $\Tr\left[\hat{P}_i\partial_\theta\hat{\rho}^\pm\right]$ is nonzero among the $i$.
So, there are exactly two members of the POM, $\hat{P}_+$ and $\hat{P}_-$, which produce signals at the pixels which vary to first order with $\theta$.
The final inequality is then saturated for
\begin{align}
    \Tr\left[\hat{P}_\pm\partial_\theta\hat{\rho}^\pm\right] = \Tr\left[\partial_\theta\hat{\rho}^\pm\right]
\end{align}
indicating that $\hat{P}_\pm$ must be the projection onto the positive or negative portion of the spectrum of $\partial_\theta\hat{\rho}$ respectively. Obviously the POVM which is optimal in the quantum case is still available, but it is not in general optimal since if $\lambda_i$ denotes the spectrum of $\partial_\theta\hat{\rho}$
\begin{align}
    \sum_{i,\lambda_i>0}\lambda_i^2\leq\left(\sum_{i, \lambda_i>0}\lambda_i\right)^2
\end{align}

As mentioned in the main text, even if $\partial_{\theta_1}\hat{\rho}$ and $\partial_{\theta_2}\hat{\rho}$ commute we may still be unable to simultaneously attain optimal measurement of both quantities.
As an example, consider the artificial family of density matrices
\begin{align}
    \hat{\rho}(t,u) = \begin{pmatrix}
        \frac{e^{9t+3u}}{1+e^{9t+3u}+e^{9t-3u}} & 0 & 0\\
        0 & \frac{e^{9t-3u}}{1+e^{9t+3u}+e^{9t-3u}} & 0\\
        0 & 0 & \frac{1}{1+e^{9t+3u}+e^{9t-3u}}
    \end{pmatrix}
\end{align}
Note that $\Tr\left[\hat{\rho}(t,u)\right] = 1$, so that while in general this formalism admits non-normalized density matrices this phenomenon does not depend on this fact.
We focus our attention on the locally optimal POM around $t=u=0$.
We find
\begin{align}
    \partial_t\hat{\rho}(0,0) = \begin{pmatrix}
        1 & 0 & 0\\
        0 & 1 & 0\\
        0 & 0 & -2
    \end{pmatrix} && \partial_u\hat{\rho}(0,0) = \begin{pmatrix}
        1 & 0 & 0\\
        0 & -1 & 0\\
        0 & 0 & 0
    \end{pmatrix}
\end{align}
which are not only simultaneously diagonalizable but already diagnoal.
The optical Cram\'er-Rao bound gives that we must have $\sigma^2\mcI_{tt}\leq 8$ and $\sigma^2\mcI_{uu}\leq 2$.
Consider two POVMs as our measurment schemes, labeled $t$ and $u$, which are optimal for their respective estimand
\begin{align}
    \hat{P}^{(t)}_1 = \begin{pmatrix}
        1 & 0 & 0\\
        0 & 1 & 0\\
        0 & 0 & 0
    \end{pmatrix} && \hat{P}_2^{(t)} = \begin{pmatrix}
        0 & 0 & 0\\
        0 & 0 & 0\\
        0 & 0 & 1
    \end{pmatrix}\\
    \hat{P}^{(u)}_1 = \begin{pmatrix}
        1 & 0 & 0\\
        0 & 0 & 0\\
        0 & 0 & 0
    \end{pmatrix} && \hat{P}^{(u)}_2 = \begin{pmatrix}
        0 & 0 & 0\\
        0 & 1 & 0\\
        0 & 0 & 0
    \end{pmatrix} && \hat{P}^{(u)}_3 = \begin{pmatrix}
        0 & 0 & 0\\
        0 & 0 & 0\\
        0 & 0 & 1
    \end{pmatrix}
\end{align}
These result in the Fisher information matrices
\begin{align}
    \sigma^2\mcI^{(t)} = \begin{pmatrix}
        8 & 0\\
        0 & 0
    \end{pmatrix} &&\sigma^2\mcI^{(u)} = \begin{pmatrix}
        6 & 0\\
        0 & 2
    \end{pmatrix}
\end{align}
demonstrating that the optimal scheme for each estimand is not optimal for the other.
Since the requirements to saturate the optical Cram\'er-Rao bound uniquely determine the required POM up to null vectors of the derivative of the density matrix, one cannot saturate both at once.

The reader may be curious how such a dramatically different form emerges in the optical case relative to the quantum one.
To illustrate this, we proceed along the standard derivation of the quantum Cram\'er-Rao bound using the form of the classical Fisher information in the optical case and note where we find an obstruction.
For this analysis, we follow Reference~\cite{braunsteinStatisticalDistanceGeometry1994}.
We can still define the symmetric logarithmic derivative $\hat{L}_\theta$ such that
\begin{align}
    \partial_\theta\hat{\rho} = \frac{1}{2}\left\{\hat{L}_\theta,\hat{\rho}\right\}
\end{align}
in terms of which we have
\begin{align}
    \mcI_{\theta} &= \frac{1}{\sigma^2}\sum_i \Tr\left[\hat{P}_i\frac{1}{2}\left\{\hat{L}_\theta,\hat{\rho}\right\}\right]^2\\
    & = \frac{1}{\sigma^2}\sum_i \left(\Re\Tr\left[\hat{P}_i\hat{L}_\theta\hat{\rho}\right]\right)^2\\
    &\leq \frac{1}{\sigma^2}\sum_i \abs{\Tr\left[\hat{P}_i\hat{L}_\theta\hat{\rho}\right]}^2\\
    &\leq \frac{1}{\sigma^2}\sum_i\Tr\left[\hat{P}_i\hat{\rho}\right]\Tr\left[\hat{P}_i\hat{L}_\theta\hat{\rho}\hat{L}_\theta\right]
\end{align}
where the last inequality used the matrix Cauchy-Schwartz inequality.
In the quantum case, the factor $\Tr\left[\hat{P}_i\hat{\rho}\right]$ cancels against an identical factor in the denominator of the classical Fisher information and we're free to resum the POVM to recieve the bound
\begin{align}
    \mcI^{\text{Q}}_\theta \leq \Tr\left[\hat{L}_\theta^2\hat{\rho}\right]
\end{align}
In our case, however, we simply recieve
\begin{align}
    \mcI_\theta \leq \frac{1}{\sigma^2}\Tr\left[\hat{\rho}^M\hat{L}_\theta\hat{\rho}\hat{L}_\theta\right]
\end{align}
where
\begin{align}
    \hat{\rho}^M = \sum_i\Tr\left[\hat{P}_i\hat{\rho}\right]\hat{P}_i
\end{align}
can be regarded as a reconstruction of the original density matrix $\hat{\rho}$ using the operators $\hat{P}_i$.
As is shown in Reference~\cite{braunsteinStatisticalDistanceGeometry1994}, saturating the matrix Cauchy-Schwartz inequality requires adapting the POVM to measure $\hat{L}_\theta$.
In our case, it is not clear from this form that this results in the maximum value for $\hat{\rho}^M$, and indeed the form of the bound and optimal POM derived above indicates that it does not.

We also note that in addition to the read noise limited case treated in the main text, if our dominant noise process is Poisson noise from $\hat{\rho}$ and the signal is large enough to approximate the distribution of this noise as Gaussian, then the classical Fisher information reads
\begin{align}
    \mcI^{L}_\theta = \sum_i \Tr\left[\hat{P}_i\partial_\theta\hat{\rho}\right]^2\left(\frac{1}{\Tr\left[\hat{P}_i\hat{\rho}\right]} + \frac{1}{2\Tr\left[\hat{P}_i\hat{\rho}\right]^2}\right)
\end{align}
where the superscipt $L$ denotes the large signal limit.
If the signal in all pixels is large enough to neglect the second summand relative to the first, we recieve
\begin{align}
    \mcI^L_\theta \approx\sum_i \frac{\Tr\left[\hat{P}_i\partial_\theta\hat{\rho}\right]^2}{\Tr\left[\hat{P}_i\hat{\rho}\right]} 
\end{align}
Thus, in the limit of very strong signals with Poisson photon statistics the optical case becomes identical to the quantum.

\section*{A distant point source seen through an aberrating medium}
Recalling the main text, the field at the aperture due to a monochomatic point source seen through a weakly aberrating medium is given by
\begin{align}
    u(\br|P,\theta) = \sqrt{\frac{P}{A}}\exp\left(i\sum_{k}\theta_k \phi_k(\br)\right)
\end{align}
where $\abs{A}$ is the aperture area and the $\phi_j$ span the phase variation over the aperture.
Note that $\norm{u(\cdot|P,\theta)}^2_\sA = P$, meaning that $P$ is the total power incident on the aperture.

We have some freedom in how we choose to parametrize the phases, so we demanded
\begin{align}\label{norm}
    \int_A\phi_j\dd \br= 0 && \innprod{\phi_j}{\phi_k}_\sA = \abs{A}\delta_{jk}
\end{align}
We can see immediately that
\begin{align}
    \innprod{u(\cdot|P,\theta)}{u(\cdot|Q,\theta^\prime)} = \sqrt{PQ}\kappa(\theta^\prime -\theta)
\end{align}
where
\begin{align}\label{kappadef}
    \kappa(\theta) = \frac{1}{\abs{A}}\int_A \exp\left(-i\sum_{j}\theta_j \phi_j(\br)\right)\dd\br
\end{align}
From this, we can deduce
\begin{align}
    \innprod{u(\cdot|P,\theta)}{\partial_{\theta_j} u(\cdot|P,\theta)} &= 0\\
    \innprod{\partial_{\theta_j} u(\cdot|P,\theta)}{\partial_{\theta_k} u(\cdot|P,\theta)} &= P\delta_{jk}
\end{align}
so that
\begin{align}\label{deriv}
    \partial_{\theta_k}\hat{\rho} = P\left[u(\cdot|1,\theta)\otimes \partial_{\theta_k}u(\cdot|1,\theta)+\partial_{\theta_k}u(\cdot|1,\theta)\otimes u(\cdot|1,\theta)\right]
\end{align}
and $u(\cdot|1,\theta)$ along with $\partial_{\theta_j}u(\cdot|1,\theta)$ form an orthonormal basis for the subspace on which $\hat{\rho}(\theta)$ and its derivatives are nonzero. Applying the bound we immediately find
\begin{align}\label{aberbound}
    \mcI_{\theta_j} \leq 2\left(\frac{P}{\sigma}\right)^2
\end{align}
for each $j$ at any $\theta$.

Because the units have been scaled via Equation~\ref{norm}, we must express the phase due to any dimensionful quantity we with to estimate in terms of our chosen $\phi_j$.
In the main text, we consider the case of estimating tip and tilt (i.e.\ angle of incidence) near normal incidence using a circular aperture.
Without loss of generality, suppose that the incident wave has $\varphi_y = 0$.
Our phase function can be written explictly as
\begin{align}
    \phi_x = \frac{4x}{D}
\end{align}
where $D$ is the aperture diameter.
Working within the small angle approximation, we have that the field at the apetrure due to a point source at infinity is
\begin{align}
    u \propto \exp\left(-\frac{2\pi i \varphi_x x}{\lambda}\right)
\end{align}
Equating this with our parametrization, we can find
\begin{align}
    \varphi_{x} = \frac{2\lambda}{\pi D}\theta_{x}
\end{align}
Applying the bound for the Fisher information given in Equation~\ref{aberbound} gives that any estimator of $\varphi_{x}$must have
\begin{align}
    \sigma_{\varphi_{x}}\geq \frac{\sqrt{2}}{\pi} \frac{\lambda}{D} \frac{\sigma}{P}
\end{align}

We also compute the resulting bound on the accuracy of depth estimation of a point source by measuring wavefront curvature using a circular aperture.
Here we have the defocus term
\begin{align}
    \phi_{020} = \sqrt{3}\left(8\frac{\br^2}{D^2} - 1\right)
\end{align}
On the other hand, in the Fresnel regime, the field at the apeture due to a point source on axis at a distance $z$ is given by
\begin{align}
    u = C\exp\left(-i\frac{\pi\br^2}{\lambda z}\right)
\end{align}
where $C$ does not depend on $\br$.
Equating the two phase factors, we find
\begin{align}
    \theta_{020} = \frac{\pi D^2}{8\sqrt{3}\lambda z}
\end{align}
which implies
\begin{align}
    \partial_{z} = -\frac{\pi D^2}{8\sqrt{3}\lambda z^2}\partial_{\theta_{020}}
\end{align}
and therefore
\begin{align}
    \sqrt{\mcI_z}=\frac{\pi D^2}{8\sqrt{3}\lambda z^2}\sqrt{\mcI_{\theta_{020}}} =\frac{\sqrt{2}\pi D^2}{8\sqrt{3}\lambda z^2}\frac{P}{\sigma}
\end{align}
So, finally applying the bound, we have
\begin{align}
    \sigma_z \geq \frac{8\sqrt{3}}{\sqrt{2}\pi} \lambda \left(\frac{z}{D}\right)^2\frac{\sigma}{P}
\end{align}

